\documentclass[10pt]{JHEP3} 

\usepackage{epsfig,multicol,bbm}

\newcommand\fverb{\setbox\fverbbox=\hbox\bgroup\verb}
\newcommand\fverbdo{\egroup\medskip\noindent%
            \fbox{\unhbox\fverbbox}\ }
\newcommand\fverbit{\egroup\item[\fbox{\unhbox\fverbbox}]}
\newbox\fverbbox

\usepackage{amsmath}
\usepackage{amssymb}
\usepackage{amsfonts}
\usepackage{latexsym}
\usepackage{graphicx}
\usepackage{dcolumn}
\usepackage{bm}
\usepackage{empheq}
\usepackage{times}  
\newcommand{\eq}[1]{\begin{equation}#1\end{equation}}

\newcommand{\ea}[1]{\begin{equation}\begin{aligned}#1\end{aligned}\end{equation}}

\newcommand{\pd}[2]{\frac{\partial #1}{\partial #2}}  
\newcommand{\lrp}[1]{\left( #1 \right)}  
\newcommand{\lrb}[1]{\left( #1 \right)}  
\newcommand{\lrsb}[1]{\left[ #1 \right]}  
\newcommand{\lrab}[1]{\left\langle #1 \right\rangle}  
\newcommand{\ab}[1]{\langle #1 \rangle} 

\def\tr{\mathrm{tr}}
\def\rd{\partial}
\def\vx{\bm{x}}
\def\vk{\bm{k}}

\def\dep{\delta\phi}
\def\dop{\dot{\phi}}

\def\qsg{Q_{\sigma}}
\def\qs{Q_{s}}

\def\doq{\dot{Q}}
\def\ca{c_{\textrm{a}}}
\def\ce{c_{\textrm{e}}}
\def\Rc{\mathcal{R}}
\def\Sc{\mathcal{S}}

\title{On the primordial trispectrum from exchanging scalar modes in general multiple field inflationary models}

\author{Xian Gao\\
    Key Laboratory of Frontiers in Theoretical Physics,\\
     Institute of Theoretical Physics, Chinese Academy of
    Sciences\\
    No.55, Zhong-Guan-Cun East Road, Hai-Dian District, Beijing 100080,
    P.R.China\\
    E-mail: \email{gaoxian@itp.ac.cn}
    }

\author{Chunshan Lin\\
    The Interdisciplinary Center for Theoretical Study,\\
     University of Science and Technology of China, Hefei, Anhui 230026, P.R.China\\
     and\\
     Department of Physics, McGill University\\
        Montr\'{e}al, QC, H3A 2T8, Canada\\
    E-mail: \email{lics@mail.ustc.edu.cn}
    }

\preprint{CAS-KITPC/ITP-194\\
            USTC/ICTS-10-09}

\abstract{ We make an complementary investigation of  the
     primordial trispectrum from exchanging intermediate scalar  modes in multi-field inflationary models with generalized kinetic terms. Together with the calculation of irreducible contributions to the primordial trispectrum in Ref.\cite{Gao:2009at}, we give the full leading-order primordial trispectrum in generalized multi-field models.
    }


\begin{document}

\section{Introduction}

One of the most exciting ideas of modern cosmology is inflation
\cite{guth1981}, which can solve the flatness, the horizon, and
the monopole problem of the standard big bang cosmology. Such a
period of cosmological inflation can be attained if the energy
density of the universe is dominated by the vacuum energy density
associated with the potential of some scalar field(s). Over the
years, inflation has become so popular because of its prediction
of nearly scale-invariant primordial density perturbation. In the
inflationary scenario, the primordial fluctuations of quantum
origin were generated and frozen to seed wrinkles in the Cosmic
Microwave Background(CMB) \cite{Miller:1999qz}\cite{de
Bernardis:2000gy}\cite{Hanany:2000qf}\cite{Halverson:2001yy}\cite{Mason:2002tm}\cite{Komatsu:2008hk}\cite{Komatsu:2010fb}\cite{Larson:2010gs}
and today's Large-scale Structure (LSS) \cite{lls
mukhanov}\cite{lls guth}\cite{lls hawking}\cite{lls
starobinsky}\cite{lls Bardeen}.


Inflation is mostly a framework of theories rather than a single
model or theory. From the observational point of view, many
inflationary models are ``degenerate". Measuring tensor modes in
the CMB anisotropy and the spectral index of the power spectrum of
adiabatic perturbation are not adequate to efficiently
discriminate among different inflationary scenarios. Fortunately,
we have another
 observable available, which proves to be valuable in providing us with additional information beyond the power spectrum to
discriminate models. It is the deviation from a purely Gaussian
statistics among CMB anisotropies \cite{komastu
0902}\cite{Bartolo:2004if}, which arises from interaction(s) among
perturbations, leading to  non-vanishing higher-order correlated
functions. Due to its importance, constraining and predicting
primordial  non-Gaussianity has become one of the major efforts in
modern cosmological community.

The simplest single-field slow-roll inflation models, within the
context of Einstein gravity and the standard initial adiabatic
vacuum, is only able to generate negligible amount of
non-Gaussianity \cite{Maldacena:2002vr}, which is undetectable by
current observations of the CMB or even LSS. In the theoretical
aspect, there are several ways to approach large non-Gaussianity.
A short list of these models and mechanisms
 includes $k$-inflation or models with  general non-canonical kinetic terms
\cite{Seery:2005wm,Chen:2006nt,Huang:2006eh,Cheung:2007st,Chen:2006xjb,Chen:2008wn,Seery:2006vu,Seery:2006js,Byrnes:2006vq,Arroja:2008ga,Seery:2008ax,Chen:2009bc,Arroja:2009pd,ArmendarizPicon:1999rj,Garriga:1999vw,Bartolo:2010bj,Bartolo:2010di},
multi-field
inflation\cite{Bartolo:2001cw,Seery:2005gb,Langlois:2008qf,Langlois:2008wt,Langlois:2009ej,RenauxPetel:2009sj,Gao:2008dt,Gao:2009gd,Arroja:2008yy,Kawakami:2009iu,Mizuno:2009cv,Mizuno:2009mv,Lehners:2009ja,Bernardeau:2002jy,Wands:2002bn,Byrnes:2008wi,Byrnes:2008zy,Battefeld:2009ym,Langlois:2008vk,Kawasaki:2008sn,Langlois:2008mn,Rigopoulos:2005xx,Ji:2009yw,Pi:2009an,Huang:2007hh,Cai:2008if,Cai:2009hw,Gao:2009qy,Wang:2010si,Sakharov:1993qh},
the curvaton scenario
\cite{Sasaki:2006kq,Malik:2006pm,Huang:2008bg,Huang:2008rj,Huang:2008ze,Li:2008fma,Ichikawa:2008iq,Kobayashi:2009cm,Li:2008jn,Gong:2009dh,Lin:2010ua,Zhang:2009gw,Cai:2010rt}
, inhomogeneous ``end-of-inflation" models such as
hybrid/multibrid models
\cite{Alabidi:2006hg,Sasaki:2008uc,Naruko:2008sq,Huang:2009vk,Huang:2009xa},
cosmic string
\cite{Regan:2009hv,Hindmarsh:2009qk,Hindmarsh:2009es}, loops
\cite{Kumar:2009ge,Cogollo:2008bi,Rodriguez:2008hy},  modified
initial vacuum \cite{Meerburg:2009ys,Meerburg:2009fi}, ghost
inflation \cite{Huang:2010ab,Izumi:2010wm}, quasi-single field
model \cite{Chen:2009we,Chen:2009zp}, vector fields
\cite{ValenzuelaToledo:2009nq,Bartolo:2009kg,Bartolo:2009pa,Karciauskas:2008bc,ValenzuelaToledo:2009af,Dimastrogiovanni:2010sm}
and so on.

Since much more observational data will be available in the near
future from WMAP/PLANCK and LSS experiments, it is very necessary
to study the four and higher-point correlation functions.  In this
paper, we make a complement to the calculation of Ref.
\cite{Gao:2009at}, in which we calculated the contributions to the
primordial trispectrum in general multi-field inflation  from the
irreducible or so-called ``contact" diagrams. A complete
calculation of the trispectrum should also include the
contributions from reducible or so-called ``exchanging
intermediate scalar modes" diagrams, as performed in
\cite{Gao:2009gd,Chen:2009bc,Arroja:2009pd,Mizuno:2009cv} in the
investigation of the trispectrum in single-field and multi-field
inflationary models, and in \cite{Seery:2008ax} where exchanging
gravitons was considered. In this paper we show that, the
contributions to the final trispectrum arising from exchanging
scalar modes has the same magnitude as those from the contact
contributions, and thus is also very important.

The remainder of this paper is organized as follows. In Sec.2, we
briefly review the background evolution and linear perturbations
for our model. Readers who are interested in the details are
encouraged to refer to \cite{Gao:2009at}. In Sec.3, we calculate
the tri-spectrum which originating from correlating (or
exchanging) scalar modes. The full trispectrum, which includes
both contacting and correlating scalar contributions, is also
discussed.







\section{Basic Setup}

\subsection{Model and Background}

In this work we consider a general class of multi-field models
containing $\mathcal{N}$ scalar fields coupled to Einstein
gravity. The action takes the form
    \eq{{\label{action_original}}
        S = \int d^4x\, \sqrt{-g} \lrsb{ \frac{R}{2} +
        P \lrp{X^{IJ},\phi^I}} \,,
        }
where $\phi^I$ ($I=1,2,\cdots,\mathcal{N}$) are scalar fields
acting as inflaton fields, and
    \eq{
        X^{IJ} \equiv -\frac{1}{2} g^{\mu\nu} \rd_{\mu} \phi^I \rd_{\nu}
        \phi^J \,,
        }
is the kinetic term (matrix), $g_{\mu\nu}$ is the spacetime metric
tensor with signature $(-,+,+,+)$. ``$I,J$"-indices are raised,
lowered and contracted by the $\mathcal{N}$-dimensional
field-space metric $G_{IJ} = G_{IJ}(\phi^I)$. This form of the
Lagrangian includes multi-field k-inflation and multi-DBI models
as special cases. For example, multi-field k-inflation has the
scalar-field Lagrangian as $P(X,\phi^I)$, where $X \equiv \tr
X^{IJ} = G_{IJ} X^{IJ}$, while in multi-field DBI models,
$P(X^{IJ},\phi^I) = - \frac{1}{f(\phi^I)} \lrp{ \sqrt{ \mathcal{D}
}-1} - V(\phi^I)$ with $ \mathcal{D} = 1 -2f G_{IJ}X^{IJ} + 4f^2
X^{[I}_I X^{J]}_J - 8f^3 X^{[I}_I X^J_J X^{K]}_K + 16 f^4 X^{[I}_I
X^J_J X^K_K X^{L]}_L$.

We work in the ADM formalism of gravitation, in which the
spacetime metric is written as
    \eq{
        ds^2 = -N^2 dt^2 +h_{ij}(dx^i+N^idt)(dx^j+N^jdt)\;,
    }
where $N=N(t,\vx)$ is the lapse function,  $N_i=N_i(t,\vx)$ is the
shift vector, and $h_{ij}$ is the spatial metric on constant time
hypersurfaces. The ADM formalism is convenient because the
equations of motion for $N$ and $N^i$ are exactly the energy and
momentum constraints which are easy to solve. Under the ADM
formalism, the action (\ref{action_original}) can be written as
(up to total derivative terms)
    \eq{{\label{action_einstein}}
    S = \int dtd^3x\, \sqrt{h} N \lrp{ \frac{1}{2}R^{(3)} + \frac{1}{2N^2} \lrp{ E_{ij}E^{ij}-E^2 } } +
    \int dtd^3x\, \sqrt{h}N\,P  \,,
    }
where $h\equiv \det h_{ij}$ and the symmetric tensor \eq{
    E_{ij} \equiv \frac{1}{2} \lrp{ \dot{h}_{ij} - \nabla_iN_j - \nabla_jN_i
    } \,,
} with $\nabla_i$ the spatial covariant derivative defined with
the spatial metric $h_{ij}$ and $E \equiv \mathrm{tr}E_{ij} =
h^{ij}E_{ij}$. $R^{(3)}$ is the three-dimensional Ricci scalar
which is computed from the spatial metric $h_{ij}$. In the ADM
formalism, spatial indices are raised and lowered using $h_{ij}$
and $h^{ij}$.

In the ADM formalism, the kinetic matrix $X^{IJ}$ can be written
as
    \eq{
        X^{IJ} = -\frac{1}{2} h^{ij} \rd_i \phi^I \rd_j \phi^J +
        \frac{1}{2N^2} v^Iv^J \,,
    }
where $v^I \equiv \dop^I - N^i \nabla_i \phi^I $.

\subsubsection{Equations of Motion}

The equations of motion for the scalar fields are
    \eq{
        \nabla_{\mu} \lrp{ P_{,\ab{IJ} } \rd^{\mu} \phi^I } + P_{,J} = 0\,,
    }
where $\nabla_{\mu}$ is the four-dimensional covariant derivative.
Here and in what follows, we denote
    \eq{
        P_{,\ab{IJ} } \equiv \pd{P}{X^{IJ}}\,,\qquad P_{,\ab{IJ}\ab{KL} }
        \equiv \frac{ \rd^2 P}{ \rd X^{IJ} \rd X^{KL}} \,,
    }
as a shorthand notation.

The equations of motion for $N$ and $N_i$ are the Hamiltonian and
momentum constraints respectively,
    \ea{{\label{constraints}}
        R^{(3)} + 2P - \frac{2}{N^2} P_{,\ab{IJ}} v^I v^J -
        \frac{1}{N^2} \lrp{ E_{ij}E^{ij} - E^2 } &=0\,,\\
        \nabla_j \lrp{ \frac{1}{N} \lrp{ E^j_i - E \delta^j_i } } -
        \frac{P_{,\ab{IJ}}}{N} v^I \nabla_i\phi^J &=0 \,.
    }

\subsubsection{Background}

In this work, we investigate scalar perturbations around a flat
FRW background, the background spacetime metric takes the form
    \eq{
        ds^2 = -dt^2 + a^2(t) \delta_{ij} dx^i dx^j \,,
    }
where $a(t)$ is the so-called scale-factor. The Friedmann equation
and the continuity equation are
    \ea{
        H^2 &= \frac{\rho}{3} \equiv \frac{1}{3} \lrp{ 2X^{IJ} P_{,\ab{IJ} } - P
        } \,,\\
        \dot{\rho} &= -3H( \rho + P ) \,.
    }
In the above equations, all quantities are background values. From
the above two equations we can also get another convenient
equation
    \eq{{\label{eq_dotH}}
        \dot{H} = -X^{IJ} P_{,\ab{IJ}} \,.
    }
The background equations of motion for the scalar fields are
    \eq{
        P_{,\ab{IJ}} \ddot{\phi}^I + \lrp{ 3HP_{,\ab{IJ}} + \dot{P}_{,\ab{IJ}}
        } \dop^I - P_{,J} = 0\,,
    }
where $P_{,I}$ denotes derivative of $P$ with respect to $\phi^I$:
$P_{,I} \equiv \pd{P}{\phi^I}$.

In this work, we investigate cosmological perturbations during an
exponential inflationary period. Thus, from (\ref{eq_dotH}) it is
convenient to define a slow-roll parameter
    \eq{
        \epsilon \equiv -\frac{\dot{H}}{H^2} = \frac{P_{,\ab{IJ}} \dop_0^I
        \dop_0^J}{2H^2} \,.
    }

\subsection{Perturbation Theory in the Spatially-flat Gauge}

The scalar metric fluctuations about our background can be written
as (see \cite{MFB,RHBrev1} for nice reviews of the theory of
cosmological perturbations) 
\ea{
    \delta N  &= \alpha \, , \\
    \delta N_i  &= \rd_i \beta \, , \\
    \delta g_{ij}  &=  - 2 a^2 \bigl( \psi \delta_{ij} - \partial_i \partial_j E) \,
    }
where $\alpha, \beta, \psi$ and $E$ are functions of space and
time{\footnote{This form of ansatz corresponds to $\delta g_{00} =
1- N^2 + N_iN^i$ and $ \delta g_{0i} = N_i $.}}.
 The scalar field perturbations are denoted by $\dep^I \equiv Q^I$.

Before proceeding, we would like to analyze the (scalar) dynamical
degrees of freedom in our system. In the beginning we have
$\mathcal{N}+4$ apparent scalar degrees of freedom. The
diffeomorphism of Einstein gravity eliminates two of
them{\footnote{See \cite{MFB} for a detailed discussion on the
gauge issue of cosmological perturbations.}}, leaving us
$\mathcal{N}+2$ scalar degrees of freedom. Furthermore, two of
these $\mathcal{N}+2$ degrees of freedom are non-dynamical. In the
ADM formalism, these are just the fluctuations $\delta N=\alpha$
and $\delta N_i = \rd_i \beta$. Thus, there are $\mathcal{N}$
propagating degrees of freedom in our system. As has been
addressed, the diffeomorphism invariance allows us to choose
convenient gauges to eliminate two degrees of freedom. In
single-field models, there are two convenient gauge choices:
comoving gauge corresponding to choosing $\dep=E=0$ or
spatially-flat gauge corresponding to $\psi=E=0$. In the
multi-field case, the comoving gauge loses its convenience since
we cannot set $\delta\rho=0$ for every field in multi-field case.
Thus, in this work we use the spatially-flat gauge.

In the spatially-flat gauge, propagating degrees of freedom for
scalar perturbations are the inflaton field perturbations
$Q^I(t,\vx)$, while  $\delta N$ and $\delta N_i$ are non-dynamical
constraints. In this work, we focus on scalar perturbations. In
general, it is well-known that in the higher-order perturbation
theories, scalar/vector/tensor perturbation modes are coupled
together. However, from the point of view of the perturbation
action approach, these couplings are equivalent to exchanging
various modes.
In this work, we focus on interactions of scalar modes themselves,
and neglect tensor perturbations. The perturbations take the form
\ea{{\label{pert_ansatz}}
    \phi^I(t,\vx) &= \phi_0^I(t) + Q^I(t,\vx) \,,\\
    h_{ij} &\equiv a^2 \delta_{ij} \,\\
    N &= 1+\alpha_1 + \alpha_2 + \cdots \,,\\
    N_i &= \rd_i (\beta_1 + \beta_2 +\cdots) + \theta_{1i} +
    \theta_{2i} +\cdots \,,
} where $\phi_0^I(t)$ is the background value, and
$\alpha_n,\beta_{n},\theta_{ni}$ are of order $\mathcal{O}(Q^n)$.

The next step is to solve the constraints $\alpha_n$, $\beta_n$
and $\theta_{ni}$ in terms of $Q^I$. Fortunately, in order to
expand the action to third-order in $Q^I$, the solutions for the
constraints up to the first-order are adequate.
 At the first-order in $Q^I$, a
particular solution for equations (\ref{constraints}) is:
    \ea{{\label{con_sol_1}}
        \alpha_1 &= \frac{1}{2H} P_{,\ab{IJ}} \dop^I Q^J \,,\\
        \beta_1 &=  \frac{a^2}{2H} \rd^{-2} \left[ \lrp{  P_{,\ab{IJ}} + 2\, X^{KL} P_{,\ab{IJ}\ab{KL}} } \lrp{ \frac{X^{IJ}}{H} P_{,\ab{KL}} \dop^K Q^L  -\dop^I \doq^J} \right. \\
        &\qquad\qquad\qquad\qquad \left. -3H P_{,\ab{IJ}} \dop^I Q^J - P_{,\ab{IJ}K}Q^K 2X^{IJ} +
        P_{,I}Q^I \right] \,,\\
        \theta_{1i} &=0 \,.
    }
Here and in what follows, repeated lower indices are contracted
using $\delta_{ij}$, and $\rd^2 \equiv \rd_i\rd_i$. $\rd^{-2}$ is
a formal notation and should be understood in fourier space.

\subsection{Linear Perturbations}
In multi-field model, we can decompose the perturbation into one
instantaneous adiabatic sector and one instantaneous entropy
sector. The ``adiabatic direction" corresponds to the direction of
the ``background inflaton velocity"
    \eq{
        e^I_1 \equiv \frac{\dop^I}{ \sqrt{ P_{,\ab{JK}} \dop^J\dop^K
        }} \equiv \frac{\dop^I}{\dot{\sigma}}\,,
    }
where we define $\dot{\sigma} \equiv \sqrt{ P_{,\ab{JK}} \dop^J\dop^K}$,
which is the generalization of the background inflaton velocity.
Actually $\dot{\sigma}$ is essentially a shorthand  notation and
has nothing to do with any concrete field. Note that
$\dot{\sigma}$ is related to the slow-roll parameter $\epsilon$ as
$\dot{\sigma}^2 = 2H^2 \epsilon$.

We introduce $(\mathcal{N}-1)$ basis $e^I_n$,
($n=2,\cdots,\mathcal{N}$) which are orthogonal with $e^I_1$ and
also with each other. The orthogonal condition can be defined as
    \eq{{\label{orthogonal}}
        P_{,\ab{IJ}} e^I_m e^J_n \equiv \delta_{mn} \,.
    }
Thus the scalar-field perturbation $Q^I$ can be decomposed into
instantaneous adiabatic/entropy basis:
    \eq{
        Q^I \equiv e^I_m Q^m \,,\qquad m=1,\cdots \mathcal{N} \,.
    }

Up to now our discussion is rather general, without further
restriction on the structure of $P(X^{IJ},\phi^I)$. In this work,
we consider a general class of two-field models, with the
following Lagrangian of the scalar fields {\footnote{ This form of
Lagrangian is motivated from that, for multi-field $k$-inflation
models \cite{Langlois:2008mn,Gao:2008dt}, the Lagrangian is simply
$P(X,\phi^I)$. In \cite{Arroja:2008yy} a special form of the
Lagrangian $ \tilde{P}(\tilde{Y},\phi^I) $ with $\tilde{Y} \equiv
X + \frac{b(\phi^I)}{2} \lrp{ X^2 - X_{IJ}X^{IJ} }$ was chosen in
the investigation of bispectrua in two-field models, which is
motivated by the multi-field DBI action. In this work, we use the
more general form of the Lagrangian (\ref{specific_model}).}}:
    \eq{{\label{specific_model}}
        P(X^{IJ},\phi^I) = P(X,Y,\phi^I) \,,
    }
with $X\equiv  X^I_I =  G_{IJ}X^{IJ}$ and $Y\equiv X^I_J X^J_I $.
This form of Lagrangian not only is the most general Lagrangian for
two-field models and thus deserves detailed investigations, but also
can make our discussions on the non-Gaussianities in two-field
models in a more general background.

After performing the decomposition into instantaneous
adiabatic/entropy modes, at the leading-order, the second-order
action for the perturbations takes the form{\footnote{In
(\ref{S2_Q}) we neglect the mass-square terms as
$\mathcal{M}_{mn}Q^m Q^n$ and the friction terms such as $\sim
\dot{Q}_m Q_n$. In general these terms may become important,
especially they may cause non-vanishing cross-correlations between
adiabatic mode and entropy mode around horizon-crossing. See
\cite{Gao:2009qy} for detailed investigation of these
cross-correlations for the same model in this paper, and
\cite{Kaiser:2010yu,Peterson:2010np} for recent studies on
multi-field perturbations.}}
 \eq{{\label{S2_Q}}
    S_2^{\textrm{(main)}} = \int dt d^3x\, a^3 \lrp{ \frac{1}{2} \mathcal{K}_{mn} \doq_m \doq_n - \frac{1}{2a^2} \delta_{mn} \rd_i Q_m \rd_i
        Q_n} \,,\\
} with \ea{{\label{K_mn}}
    \mathcal{K}_{mn} &\equiv \delta_{mn} + \lrp{P_{,\ab{MN}} \dop^M
        \dop^N} P_{,\ab{IK} \ab{JL}} e^I_1 e^K_n e^J_1 e^L_m \,,\\
        &= \delta_{mn} + \lrp{ \frac{1}{\ca^2}-1 }
        \delta_{1m} \delta_{1n} + \lrp{ \frac{1}{\ce^2} - 1 }
        \lrp{\delta_{mn}- \delta_{1m} \delta_{1n}
        } \,,
        }
where we introduce{\footnote{We use $\ca$ and $\ce$ rather than
$c_{\sigma}$ and $c_s$ in order to avoid possible confusion, since
in the literatures $c_s$ has special meaning, i.e. the speed of
sound of perturbation in single-field models.}}
    \ea{{\label{def_ca_ce}}
        \ca^2 &\equiv \frac{P_{,X}+2XP_{,Y}}{P_{,X}+2X\left(P_{,XX}+4XP_{,XY}+3P_{,Y}+4X^{2}P_{,YY}\right)} \,,\\
        \ce^2 &\equiv \frac{P_{,X}}{P_{,X}+2XP_{,Y}} \,,
    }
which are the propagation speeds of adiabatic and entropy
perturbations respectively. It is useful to note that
$\mathcal{K}_{mn}$ is diagonal, $\mathcal{K}_{11} = 1/\ca^2$,
$\mathcal{K}_{22} = 1/\ce^2$ and $\mathcal{K}_{12} =
\mathcal{K}_{21} = 0$, as a consequence of the adiabatic/entropy
decomposition. $\ca \neq \ce$ is a generic feature in multi-field
models; this can be seen explicitly from the definitions in
(\ref{def_ca_ce}), the speed of sound for the adiabatic mode and
the entropy mode(s) have different dependence on the
$P$-derivatives\footnote{This fact was first point out apparently
in \cite{Easson:2007dh,Huang:2007hh} in the investigation of brane
inflation models. See also
\cite{Arroja:2008yy,Langlois:2008qf,Cai:2008if,Cai:2009hw,Ji:2009yw,Gao:2008dt,Gao:2009qy}
for extensive investigations on general multi-field models with
different $\ca$ and $\ce$.}.

At this point, it is convenient to introduce two parameters:
    \ea{{\label{def_lambda}}
    \xi&\equiv
    \frac{X(P_{,XX}+2P_{,XY})}{P_{,X}+2XP_{,Y}}~,\nonumber\\
        \lambda &\equiv X^{2}P_{,XX}+\frac{2}{3}X^{3}P_{,XXX}+2\left(YP_{Y}+6Y^{2}P_{,YY}+\frac{8}{3}Y^{3}P_{,YYY}\right) \\
        &\qquad +4\left(X^{2}YP_{,XXY}+2XYP_{,XY}+2XY^{2}P_{,XYY}\right)
        \,,
    }
where all quantities are background values, and we have used
$Y=X^2$. As we will see later, although the $X$,$Y$-dependences of
$P(X,Y,\phi^I)$ in general can be complicated, the non-linear
structures of $P$ affect the trispectra through the above specific
combinations of derivatives of $P$.

After introducing new variables whose kinetic terms are canonically normalized
    \eq{
        \tilde{Q}_{\sigma} \equiv  \frac{a}{\ca} \qsg \,,\qquad\qquad \tilde{Q}_{s}
        \equiv \frac{a}{\ce} \qs \,,
    }
and changing
into comoving time defined by $dt=a d\eta$, the quadratic action
takes the form
    \eq{{\label{S2_tilde_Q}}
        S_2 = \int d\eta d^3x\,\frac{1}{2} \lrsb{ \tilde{Q}'^2_{\sigma} + \lrp{ \mathcal{H}^2 + \mathcal{H}' } \tilde{Q}_{\sigma}^2 - \ca^2 (\rd \tilde{Q}_{\sigma})^2 + \tilde{Q}'^2_{s} + \lrp{ \mathcal{H}^2 + \mathcal{H}' } \tilde{Q}_{s}^2 - \ce^2 (\rd \tilde{Q}_{s})^2   }\,.
    }
The action (\ref{S2_Q}) or (\ref{S2_tilde_Q}) describes a free
theory. Performing a canonical quantization, we write
    \eq{
        \tilde{Q}_{\sigma}(\vk,\eta) \equiv a_{\vk} \tilde{u}_k(\eta) + a^{\dag}_{-\vk}
        \tilde{u}^{\ast}_k(\eta) \,, \qquad\qquad \tilde{Q}_s(\vk,\eta) \equiv a_{\vk} \tilde{v}_k(\eta) + a^{\dag}_{-\vk}
        \tilde{v}^{\ast}_k(\eta) \,,
    }
where $\tilde{u}_k(\eta)$ and $\tilde{v}_k(\eta)$ are the mode
functions, which satisfy the corresponding classical equations of
motion
    \ea{{\label{mode_function_tilde}}
        \tilde{u}''_k + \lrsb{ \ca^2 k^2 - ( \mathcal{H}^2 + \mathcal{H}' ) } \tilde{u}_k =0   \,,\qquad\qquad \tilde{v}''_k + \lrsb{ \ce^2 k^2 - ( \mathcal{H}^2 + \mathcal{H}' )
}  \tilde{v}_k  =0  \,.
    }

Finally, what we are interested in are the tree-level two-point
functions for $\qsg$ and $\qs$, defined as
    \ea{
        \lrab{ \qsg(\vk_1,\eta_{1}) \qsg(\vk_2,\eta_{2}) } = (2\pi)^3 \delta^2 (\vk_1 + \vk_2)
        G_{k_1}(\eta_1,\eta_2) \,,\\
        \lrab{ \qs(\vk_1,\eta_{1}) \qs(\vk_2,\eta_{2}) } = (2\pi)^3 \delta^2 (\vk_1 + \vk_2)
        F_{k_1}(\eta_1,\eta_2) \,,\\
    }
with
    \eq{
        G_k(\eta_1,\eta_2) \equiv u_k(\eta_1) u^{\ast}_k(\eta_2)
        \,,\qquad \qquad F_k(\eta_1,\eta_2) \equiv v_k(\eta_1)
        v^{\ast}_k(\eta_2) \,,
    }
where $u_k(\eta)$ and $v_k(\eta)$ are the mode functions for
adiabatic perturbation and entropy perturbation respectively:
    \ea{
        u_k(\eta) &= \frac{i\, H}{ \sqrt{ 2\ca k^3 }} \lrp{ 1+ i\ca k \eta } e^{-i \ca k \eta}  \,,\\
        v_k(\eta) &=  \frac{i\, H}{ \sqrt{ 2\ce k^3 }} \lrp{ 1+ i\ce k \eta } e^{-i \ce k \eta} \,.
    }

The so-called ``power spectra" for adiabatic and entropy
perturbations are defined as $P_{\sigma}(k) \equiv
G_k(\eta_{\ast},\eta_{\ast})$ and $P_{s}(k) \equiv
F_k(\eta_{\ast},\eta_{\ast})$, where $\eta_{\ast}$ can be chosen
as the time when the modes cross the sound-horizon, i.e. at $\ca
k\equiv aH$ for adiabatic mode and $\ce k \equiv aH$ for entropy
mode(s){\footnote{In general multi-field models,
adiabatic/entropic modess with the same comoving wavenumber $k$
exit their sound-horizons at different time, due to their
different speeds of sound, $\ca \neq \ce$. This fact will bring
new interesting phenomenology in multi-field models.
 As was shown in \cite{Gao:2009qy}, the cross-correlations between adiabatic/entropic modes would be enhanced
 by a small $\ce/\ca$ ratio.}}. 
In the so-called comoving gauge, the perturbation $\qsg$ is
directly related to the three-dimensional curvature of constant
time space-like slices. This gives the gauge-invariant quantity
referred to as the ``comoving curvature perturbation":
    \eq{
        \mathcal{R} \equiv \frac{H}{\dot{\sigma}} \qsg \,.
    }
 The entropy
perturbation $\qs$ is automatically gauge-invariant by
construction. It is also convenient to introduce a renormalized
``isocurvature perturbation" defined as
    \eq{
        \mathcal{S} \equiv \frac{H}{\dot{\sigma}} \qs \,.
    }

In the cosmological context, it is also convenient to define the
dimensionless power spectra for comoving curvature perturbation
and isocurvature perturbation respectively:
    \ea{{\label{dimless_spec}}
        \mathcal{P}_{\mathcal{R}\ast} &= \frac{H^2}{\dot{\sigma}^2} \mathcal{P}_{\sigma\ast} \equiv \frac{H^2}{\dot{\sigma}^2} \frac{k^3}{2\pi^2} P_{\sigma\ast}(k)
        = \frac{1}{2\epsilon\ca} \lrp{ \frac{H}{2\pi} }^2\,,\\
        \mathcal{P}_{\mathcal{S}\ast} &= \frac{H^2}{\dot{\sigma}^2} \mathcal{P}_{s\ast} \equiv \frac{H^2}{\dot{\sigma}^2} \frac{k^3}{2\pi^2}  P_{s\ast}(k)
        = \frac{1}{2\epsilon\ce} \lrp{ \frac{H}{2\pi} }^2 \,.
    }
In the above results, all quantities are evaluated around the
sound-horizon crossing.

\section{Non-linear perturbations}
In this section, We calculate the tri-spectrum which comes from
correlating (or exchanging) scalar modes. The full trispectrum
which includes both contacting and correlating scalar
contributions is also discussed in this section.

\subsection{Trispectra from Correlating Scalar Mode}

The third-order action for the model (\ref{action_original}) has been derived in \cite{Arroja:2008yy}:
    \ea{{\label{action_2nd_3rd}}
        S_3^{\textrm{(main)}} &= \int dt d^3x\, a^3 \lrp{ \frac{1}{2} \Xi_{mnl} \doq_m \doq_n \doq_l
         - \frac{1}{2a^2} \Upsilon_{mnl}\, \doq_m \rd_iQ_n \rd_iQ_l } \,,
    }
with
    \ea{
        \Xi_{mnl} &\equiv \sqrt{ P_{,\ab{MN}} \dop^M\dop^N
        } \lrsb{ P_{,\ab{IK} \ab{JL}}  e^{I}_1 e^{K}_{m} e^J_{n} e^L_{\ell} +\frac{1}{3} \lrp{ P_{,\ab{MN}} \dop^M\dop^N } P_{,\ab{IK} \ab{JL} \ab{PQ}
        } e^{I}_1 e^{K}_m e^{J}_1 e^{L}_n e^{P}_1 e^{Q}_{l} }
        \,,\\
        \Upsilon_{mnl} &\equiv \sqrt{P_{,\ab{MN}}
        \dop^M\dop^N} \, P_{,\ab{IK} \ab{JL}}\,  e^{I}_1 e^{K}_m e^{J}_n e^L_l \,.
    }
In this article, we still work on the double-field model. It is a
straightforward task to generalize our calculation to a more
general multi field model.

Direct algebra gives the cubic-order interaction Hamiltonian:
    \ea{{\label{HI}}
        H_I(\tau) =&\int d\tau d^{3}x \left[-\frac{a}{2}\Xi_{\sigma}Q_{\sigma}'^{3}
        +\frac{a}{2}\Upsilon_{\sigma}Q_{\sigma}'\partial_{i}Q_{\sigma}\partial_{i}Q_{\sigma}\right.\\
        & \qquad\qquad\quad \left.-\frac{a}{2}\Xi_{c}Q_{\sigma}'Q_{s}'^{2}+\frac{a}{2}\Upsilon_{s}Q_{\sigma}'\partial_{i}
        Q_{s}\partial_{i}Q_{s}+\frac{a}{2}\Upsilon_{c}Q_{s}'\partial_{i}Q_{\sigma}\partial_{i}Q_{s}\right]
\,,
    }
where the subscript ``I" denotes the interactional picture and the
five effective couplings $\Xi_{\sigma}$ etc. are given in Appendix
A.

The trispectrum is the four-point correlation function of
perturbations. According to the in-in formalism
\cite{Weinberg:2005vy}, The trispectrum which comes from scalar
exchanging can be formulated as
\begin{eqnarray} \langle
Q^4\rangle
&\supset&-2\Re\left[\int_{-\infty^+}^{\eta}d\eta'\int_{-\infty^+}^{\eta'}d\eta''
\langle0_I|Q_I^4H_I(\eta')H_I(\eta'')|0_I\rangle\right]\nonumber\\
&&+\int_{-\infty^-}^{\eta}d\eta'
\int_{-\infty^+}^{\eta}d\eta''\langle0_I|H_I(\eta')Q_I^4H_I(\eta'')|0_I\rangle.\end{eqnarray}

The calculation is straightforward but rather tedious. Here we simply collect the final results.
 The leading contribution from exchanging an intermediate
scalar mode to the purely adiabatic four-point function
$\lrab{Q_{\sigma}^4}$ is given by (see Appendix \ref{appsec:int}
for details)
    \ea{{\label{4sigma_SE}}
        &\left\langle Q_{\sigma}\left(\tau,\bm{k}_{1}\right)Q_{\sigma}\left(\tau,\bm{k}_{2}\right)Q_{\sigma}\left(\tau,\bm{k}_{3}
        \right)Q_{\sigma}\left(\tau,\bm{k}_{4}\right)\right\rangle
        _{\textrm{SE}}\\=& (2\pi)^3\delta^{3}(\sum_i^4\vk_i) \Big\{\frac{9}{2}\Xi_{\sigma}^{2}c_{a}^{10}\mathcal{I}_{a}\left(c_{a}k_{1},c_{a}k_{2},c_{a}k_{3},c_{a}k_{4},c_{a}k_{12}\right)\\&+2\Upsilon_{\sigma}^{2}c_{a}^{6}\left[\frac{1}{4}\mathcal{I}_{b}^{\left(1\right)}\left(c_{a}k_{1},c_{a}k_{2},c_{a}k_{3},c_{a}k_{4},c_{a}k_{12}\right)+\mathcal{I}_{b}^{\left(2\right)}\left(c_{a}k_{1},c_{a}k_{2},c_{a}k_{3},c_{a}k_{4},c_{a}k_{12}\right)+\mathcal{I}_{b}^{\left(3\right)}\left(c_{a}k_{1},c_{a}k_{2},c_{a}k_{3},c_{a}k_{4},c_{a}k_{12}\right)\right]\\&-3\Xi_{\sigma}\Upsilon_{\sigma}c_{a}^{8}\left[\mathcal{I}_{c}^{\left(1\right)}\left(c_{a}k_{1},c_{a}k_{2},c_{a}k_{3},c_{a}k_{4},c_{a}k_{12}\right)+2\mathcal{I}_{c}^{\left(2\right)}\left(c_{a}k_{1},c_{a}k_{2},c_{a}k_{3},c_{a}k_{4},c_{a}k_{12}\right)\right]+23\textrm{perms}\Big\}
        \,,}
where ``23 perms" denotes the other 23 permutations among four
external momenta $\bm{k}_1 , \cdots, \bm{k}_4$. In
(\ref{4sigma_SE}), the integrals $\mathcal{I}_a$,
$\mathcal{I}_{b}^{(1)}$ etc are defined in Appendix
\ref{appsec:int}. The mixed adiabatic/entropy four-point function
$\lrab{Q_{\sigma}^2 Q_s^2}$ is given by
    \ea{{\label{2sigma2s_SE}}
        &\left\langle Q_{\sigma}\left(\tau,\bm{k}_{1}\right)Q_{\sigma}
        \left(\tau,\bm{k}_{2}\right)Q_{s}\left(\tau,\bm{k}_{3}\right)Q_{s}
        \left(\tau,\bm{k}_{4}\right)\right\rangle _{\textrm{SE}} + 5\textrm{perms}\\
        = &(2\pi)^3\delta^3(\sum_i^4\vk_i) \Big\{3\Xi_{\sigma}\Xi_{c}c_{a}^{6}c_{e}^{4}\mathcal{I}_{a}\left(c_{a}k_{1},c_{a}k_{2},c_{e}k_{3},c_{e}k_{4},c_{a}k_{12}\right)+2\Xi_{c}^{2}c_{a}^{4}c_{e}^{6}\mathcal{I}_{a}\left(c_{a}k_{1},c_{e}k_{3},c_{a}k_{2},c_{e}k_{4},c_{e}k_{13}\right)\\&+\Upsilon_{\sigma}\Upsilon_{s}c_{a}^{4}c_{e}^{2}\left[\mathcal{I}_{b}^{\left(1\right)}\left(c_{a}k_{1},c_{a}k_{2},c_{e}k_{3},c_{e}k_{4},c_{a}k_{12}\right)+2\mathcal{I}_{b}^{\left(2\right)}\left(c_{e}k_{3},c_{e}k_{4},c_{a}k_{1},c_{a}k_{2},c_{a}k_{34}\right)\right]\\&+\Upsilon_{\sigma}\Upsilon_{c}c_{a}^{3}c_{e}^{3}\left[\mathcal{I}_{b}^{\left(2\right)}\left(c_{a}k_{1},c_{a}k_{2},c_{e}k_{3},c_{e}k_{4},c_{a}k_{12}\right)+2\mathcal{I}_{b}^{\left(3\right)}\left(c_{a}k_{1},c_{a}k_{2},c_{e}k_{3},c_{e}k_{4},c_{a}k_{12}\right)\right]\\&+2\Upsilon_{s}^{2}c_{a}^{4}c_{e}^{2}\mathcal{I}_{b}^{\left(3\right)}\left(c_{a}k_{1},c_{e}k_{3},c_{a}k_{2},c_{e}k_{4},c_{e}k_{13}\right)\\&+2\Upsilon_{s}\Upsilon_{c}c_{a}^{3}c_{e}^{3}\left[\mathcal{I}_{b}^{\left(2\right)}\left(c_{a}k_{2},c_{e}k_{4},c_{a}k_{1},c_{e}k_{3},c_{e}k_{24}\right)+\mathcal{I}_{b}^{\left(3\right)}\left(c_{a}k_{1},c_{e}k_{3},c_{e}k_{4},c_{a}k_{2},c_{e}k_{13}\right)\right]\\&+\frac{1}{2}\Upsilon_{c}^{2}c_{e}^{4}c_{a}^{2}\left[\mathcal{I}_{b}^{\left(1\right)}\left(c_{a}k_{1},c_{e}k_{3},c_{a}k_{2},c_{e}k_{4},c_{e}k_{12}\right)+2\mathcal{I}_{b}^{\left(2\right)}\left(c_{a}k_{1},c_{e}k_{3},c_{e}k_{4},c_{a}k_{2},c_{e}k_{13}\right)+\mathcal{I}_{b}^{\left(3\right)}\left(c_{e}k_{3},c_{a}k_{1},c_{e}k_{4},c_{a}k_{2},c_{e}k_{13}\right)\right]\\&-3\Xi_{\sigma}\Upsilon_{s}c_{a}^{6}c_{e}^{2}\mathcal{I}_{c}^{\left(1\right)}\left(c_{a}k_{1},c_{a}k_{2},c_{e}k_{3},c_{e}k_{4},c_{a}k_{12}\right)-3\Xi_{\sigma}\Upsilon_{c}c_{a}^{5}c_{e}^{3}\mathcal{I}_{c}^{\left(2\right)}\left(c_{a}k_{1},c_{a}k_{2},c_{e}k_{3},c_{e}k_{4},c_{a}k_{12}\right)\\&-\Xi_{c}\Upsilon_{\sigma}c_{e}^{4}c_{a}^{4}\mathcal{I}_{c}^{\left(1\right)}\left(c_{e}k_{3},c_{e}k_{4},c_{a}k_{1},c_{a}k_{2},c_{a}k_{34}\right)-2\Xi_{c}\Upsilon_{\sigma}c_{e}^{4}c_{a}^{4}\mathcal{I}_{c}^{\left(2\right)}\left(c_{e}k_{3},c_{e}k_{4},c_{a}k_{1},c_{a}k_{2},c_{a}k_{34}\right)\\&-4\Xi_{c}\Upsilon_{s}c_{e}^{4}c_{a}^{4}\mathcal{I}_{c}^{\left(2\right)}\left(c_{a}k_{1},c_{e}k_{3},c_{a}k_{2},c_{e}k_{4},c_{e}k_{13}\right)\\&-2\Xi_{c}\Upsilon_{c}c_{a}^{3}c_{e}^{5}\left[\mathcal{I}_{c}^{\left(1\right)}\left(c_{a}k_{1},c_{e}k_{3},c_{a}k_{2},c_{e}k_{4},c_{e}k_{13}\right)+\mathcal{I}_{c}^{\left(2\right)}\left(c_{a}k_{1},c_{e}k_{3},c_{e}k_{4},c_{a}k_{2},c_{e}k_{13}\right)\right] + 23\textrm{perms} \Big\}\,,
    }
where in the first line ``5 perms" denotes the other 5
possibilities of choosing two momenta for $\qsg$ and two momenta
for $\qs$ out of the
 four external momenta. Note that in the permutations, the speeds of sound $\ca$ and $\ce$ are always associated with the given extra momenta. The purely entropic four-point function $\lrab{Q_{s}^4}$ is
    \ea{{\label{4s_SE}}
        &\left\langle Q_{s}\left(\tau,\bm{k}_{1}\right)Q_{s}\left(\tau,\bm{k}_{2}\right)Q_{s}\left(\tau,\bm{k}_{3}\right)Q_{s}\left(\tau,\bm{k}_{4}\right)\right\rangle
        _{\textrm{SE}}\\=& (2\pi)^3\delta^3(\sum_i^4\vk_i) \Big\{\frac{1}{2}\Xi_{c}^{2}c_{e}^{8}c_{a}^{2}\mathcal{I}_{a}
        \left(c_{e}k_{1},c_{e}k_{2},c_{e}k_{3},c_{e}k_{4},c_{a}k_{12}\right)+
        \frac{1}{2}\Upsilon_{s}^{2}c_{e}^{4}c_{a}^{2}\mathcal{I}_{b}^{\left(1\right)}
        \left(c_{e}\bm{k}_{1},c_{e}\bm{k}_{2},c_{e}\bm{k}_{3},c_{e}\bm{k}_{4},c_{a}k_{12}\right)\\
        &+\Upsilon_{s}\Upsilon_{c}c_{e}^{5}c_{a}\mathcal{I}_{b}^{\left(2\right)}\left(c_{e}\bm{k}_{1}
        ,c_{e}\bm{k}_{2},c_{e}k_{3},c_{e}\bm{k}_{4},c_{a}\bm{k}_{12}\right)+\frac{1}{2}
        \Upsilon_{c}^{2}c_{e}^{6}\mathcal{I}_{b}^{\left(3\right)}\left(c_{e}k_{1},c_{e}\bm{k}_{2},c_{e}k_{3},c_{e}
        \bm{k}_{4},c_{a}\bm{k}_{12}\right)\\&-\Xi_{c}\Upsilon_{s}c_{e}^{6}c_{a}^{2}\mathcal{I}_{c}^{\left(1\right)}
        \left(c_{e}k_{1},c_{e}k_{2},c_{e}\bm{k}_{3},c_{e}\bm{k}_{4},c_{a}k_{12}\right)-\Xi_{c}\Upsilon_{c}c_{e}^{7}
        c_{a}\mathcal{I}_{c}^{\left(2\right)}\left(c_{e}k_{1},c_{e}k_{2},c_{e}k_{3},c_{e}\bm{k}_{4},c_{a}
        \bm{k}_{12}\right)+23\textrm{perms}\Big\}
        \,.
    }

\subsection{Full Trispectrum for the Curvature Perturbation}

The scalar field perturbations $Q_{\sigma}$ and $Q_s$ themselves
are not directly observable. What we are eventually interested in
is the curvature perturbation $\mathcal{R}$. As has been
investigated in
\cite{Langlois:2008qf,Langlois:2008wt,Arroja:2008yy}, the comoving
curvature perturbation $\mathcal{R}$ is related to the adiabatic
and entropy perturbations of the scalar fields by
    \ea{
        \mathcal{R} &\approx \mathcal{R}_{\ast} +
        T_{\mathcal{R}\mathcal{S}} \mathcal{S}_{\ast} =
        \lrp{\frac{H}{\dot{\sigma}}}_{\ast} Q_{\sigma\ast} +
        T_{\mathcal{R}\mathcal{S}}\lrp{\frac{H}{\dot{\sigma}}}_{\ast}
        Q_{s\ast} \\
        &\equiv \mathcal{N}_{\sigma} Q_{\sigma\ast} + \mathcal{N}_s
        Q_{s\ast} \,.
    }
Here $T_{\mathcal{R}\mathcal{S}}$ is the so-called transfer function from entropy perturbation to adiabatic perturbation{\footnote{As was pointed in \cite{Battefeld:2006sz,Battefeld:2007en}, as long as the fields roll slowly, these additional
contributions after horizon-crossing are heavily suppressed.}}. Note that $\mathcal{N}_s \equiv T_{\mathcal{R}\mathcal{S}}
\mathcal{N}_{\sigma}$ is in general time-dependent. Thus
contributions to the four-point correlation function for
$\mathcal{R}$ 
are given by
    \ea{{\label{4R_original}}
        &\quad\; \lrab{ \Rc(\vk_1)\Rc(\vk_3)\Rc(\vk_2)\Rc(\vk_4) } \\
        &=
        \mathcal{N}_{\sigma}^4 \left[\lrab{ \qsg(\vk_1)\qsg(\vk_2)\qsg(\vk_3)\qsg(\vk_4)
        }_{\textrm{SE}}+\lrab{ \qsg(\vk_1)\qsg(\vk_2)\qsg(\vk_3)\qsg(\vk_4)
        }_{\textrm{C}} \right]\\
        &+ \mathcal{N}_{\sigma}^2 \mathcal{N}_s^2
        \lrsb{ \lrab{ \qsg(\vk_1)\qsg(\vk_2)\qs(\vk_3)\qs(\vk_4) }_{\textrm{SE}} + \lrab{ \qsg(\vk_1)\qsg(\vk_2)\qs(\vk_3)\qs(\vk_4) }_{\textrm{C}}+\textrm{5 perms} }  \\
        &\qquad\qquad + \mathcal{N}_{s}^4 \lrsb{\lrab{ \qs(\vk_1)\qs(\vk_2)\qs(\vk_3)\qs(\vk_4)
        }_{\textrm{SE}}+\lrab{ \qs(\vk_1)\qs(\vk_2)\qs(\vk_3)\qs(\vk_4)
        }_{\textrm{C}}} \,,
    }
where the subscript $_{SE}$ denote the four-point functions which
come from exchanging intermediate scalar modes, and the subscript
$_{C}$ denotes the four-point functions which come from contact
diagrams. The four-point functions of exchanging scalar modes for
$Q_{\sigma}$ and $Q_s$ are given by (\ref{4sigma_SE}),
(\ref{2sigma2s_SE}) and (\ref{4s_SE}). The four-point function of
the contact diagram is given in Ref.\cite{Gao:2009at}. In deriving
(\ref{4R_original}), we have used the assumption that there is no
cross-correlation between adiabatic and entropy modes, i.e.
$\lrab{\qsg\qs}_{\ast}\equiv 0$, around horizon-crossing.

It is convenient to define a so-called trispectrum
\begin{eqnarray}\label{socalledtrispectrum}
\langle\mathcal{R}(\vk_1)\mathcal{R}(\vk_2)\mathcal{R}(\vk_3)\mathcal{R}(\vk_4)\rangle&\equiv&(2\pi)^3\delta^3(\sum_{i=1}^{4}\vk_i)\mathcal{T}(\vk_1,\vk_2,\vk_3,\vk_4)\nonumber\\
&=&(2\pi)^3\delta^3(\sum_{i=1}^{4}\vk_i)(\mathcal{T}_{c}(\vk_1,\vk_2,\vk_3,\vk_4)+\mathcal{T}_s(\vk_1,\vk_2,\vk_3,\vk_4))
\end{eqnarray}
where $\mathcal{T}_{c}$ is given by eq.(5.29) of
Ref.\cite{Gao:2009at}. We can derive $\mathcal{T}_s$ from
eqs.(\ref{4sigma_SE})(\ref{2sigma2s_SE})(\ref{4s_SE}).



To investigate the size of Non-Gaussianity roughly, we choose
regular tetrahedron limit, $k_1=k_2=k_3=k_4=k_{12}=k_{34}$, and
take the approximation $c_a^2=c_e^2\equiv c_s^2\ll 1$. We define a
real number $t_{NL}$ from the trispectrum to characterize its
size,
\begin{eqnarray}
\mathcal{T}(k_1,k_2,k_3,k_4)|_{rth}\equiv P_{\mathcal{R}}^3
t_{NL}~.
\end{eqnarray}
We have
\begin{eqnarray}
t_{NL}=t_{NL}^c+t_{Nl}^s~,
\end{eqnarray}
where $t_{NL}^c$ comes from the contact diagram \cite{Gao:2009at},

    \ea{
         t_{\textrm{NL}}^{\textrm{c}} &= \lrp{1 + T^2_{\Rc\Sc} }^{-3} \lrp{t_1 + t_2 + t_3 }\,,
    }
where
    \ea{
        t_1 &= - \frac{3 c_s^2 \left(54 c_s^2 \lambda ^2-H^2 \epsilon  (3 \lambda +10 \Pi )\right)}{512 H^4 \epsilon ^2} - T^2_{\Rc\Sc} \frac{9 c_s \left(H^2 \epsilon -15c_s^6 \lambda \right)}{256c_s^5 H^2 \epsilon } + T^4_{\Rc\Sc} \frac{81}{1024 c_s^2}\,,\\
        t_2 &= \frac{13 \left(-H^2 \epsilon + 3 c_s^4 \lambda \right)}{256 c_s^2 H^2 \epsilon} + T^2_{\Rc\Sc}\frac{13(-H^2\epsilon+3c_s^4\lambda)}{128c_s^4H^2\epsilon}+ T^4_{\Rc\Sc}\frac{13}{256 c_s^4} \,,\\
        t_3 &=  \frac{515}{8192 c_s^2}+
        T^2_{\Rc\Sc}\frac{103}{2048c_s^2}
        \,.
    }
$t_{NL}^s$ comes from scalar exchange diagram,
\begin{eqnarray}
t_{NL}^s=t_4+t_5+t_6~,
\end{eqnarray}
where
\begin{eqnarray}\label{rthSE}
t_4&\simeq&
 2c_{s}^{4}\left(\frac{\lambda}{H^{2}\epsilon}\right)^{2}+\left(0.22c_{s}^{-4}+\frac{0.67\lambda}{H^{2}\epsilon}\right)T_{RS}^{2}+0.06c_{s}^{-4}T_{RS}^{4}~,\nonumber\\
t_5&\simeq&
2.74c_{s}^{-4}+\left(8.53+12.95\xi+5.75\xi^{2}\right)c_{s}^{-4}T_{RS}^{2}+\left(1.43+1.99\xi+1.24\xi^{2}\right)c_{s}^{-4}T_{RS}^{4}~,\nonumber\\
t_6&\simeq&
-2.25\frac{\lambda}{H^{2}\epsilon}+\left(20.61c_{s}^{-4}+14.72c_{s}^{-4}\xi+2.21\frac{\lambda}{H^{2}\epsilon}+2.30\xi\frac{\lambda}{H^{2}\epsilon}\right)T_{RS}^{2}+\left(0.37+0.38\xi\right)c_{s}^{-4}T_{RS}^{4}~,
\end{eqnarray}
Here the contributions $t_4$, $t_5$, $t_6$ come from diagram
$\mathcal{I}_a$, $\mathcal{I}_b$ and $\mathcal{I}_c$ respectively
(see Appendix \ref{appsec:int} for details). Comparing $t_4$,
$t_5$, $t_6$ with $t_1$, $t_2$, $t_3$, we can see that the scalar
exchange diagram makes a nontrivial contribution to the
trispectrum. As for the contact contributions, the contributions
to the trispectrum from exchanging scalar modes can be enhanced by
small sound speed(s), large $T_{\mathcal{RS}}$, large $\xi$, and
large $\frac{\lambda}{H^{2}\epsilon}$.

\section{Conclusion}

In this note, we made a complementary calculation of the
contributions to the trispectrum of primordial curvature
perturbations from exchanging intermediate scalar modes in the
context of generalized multi-field inflation, which completes the
calculation of our previous investigation \cite{Gao:2009at}. We
choose regular tetrahedron limit to estimate the size of
non-Gaussianity.
The calculation presented in this work, together with
\cite{Gao:2009at}, can be employed as the starting point for
further analysis of the trispectrum of generalized multi-field
inflatioyn models, such as the shapes, squeezed limit
\cite{Creminelli:2004yq,Ganc:2010ff,RenauxPetel:2010ty} and
estimators
\cite{Fergusson:2009nv,Munshi:2009wy,Mizuno:2010by,Regan:2010cn,Smidt:2010ra}
etc. We would like to come back to these issues in the near
future.



\acknowledgments

We would like to thank Miao Li, Yi Wang, Shinji Mukohyama for
useful discussion, and thank Thorsten Battefeld for the careful
reading of the manuscript and suggestions on improvement. XG is
deeply grateful to Prof. Miao Li for his consistent encouragement
and support. This work was partly supported by the NSFC grant
No.10535060/A050207,
 a NSFC group grant No.10821504 and Ministry of Science and
Technology 973 program under grant No.2007CB815401, and in part by Perimeter Institute for Theoretical Physics.
CL is
supported by Chinese Scholarship Council.
CL would like to thank the hospitality of Perimeter Institute, where the paper was finalized when CL visited PI.

\appendix

\section{Coefficients in the interactional Hamiltonian }

The variously introduced coefficients in (\ref{HI}) are given by
\begin{eqnarray}
\Xi_{\sigma}&=&\frac{4\lambda}{\dot{\sigma}^3}~,\nonumber\\
\Xi_{c}&=&\frac{H\sqrt{\epsilon}}{\sqrt{2}XP_{,X}}\left(\frac{1}{c_a^2-1}\right)~,\nonumber\\
\Upsilon_{\sigma}&=&\frac{1}{H\sqrt{2\epsilon}}\left(\frac{1}{c_a^2-1}\right)~,\nonumber\\
\Upsilon_{s}&=&\frac{\dot{\sigma}\xi}{XP_{,X}}\nonumber\\
\Upsilon_{c}&=&\frac{\sqrt{2}}{H\sqrt{\epsilon}}\left(\frac{1}{c_e^2-1}\right)~.
\end{eqnarray}

\section{Basic Integrals}{\label{appsec:int}}

\begin{figure}[h]
\centering
\begin{minipage}{0.8\textwidth}
\centering
\begin{tabular}{cc}
\begin{tabular}{c}
\includegraphics[width=4.5cm]{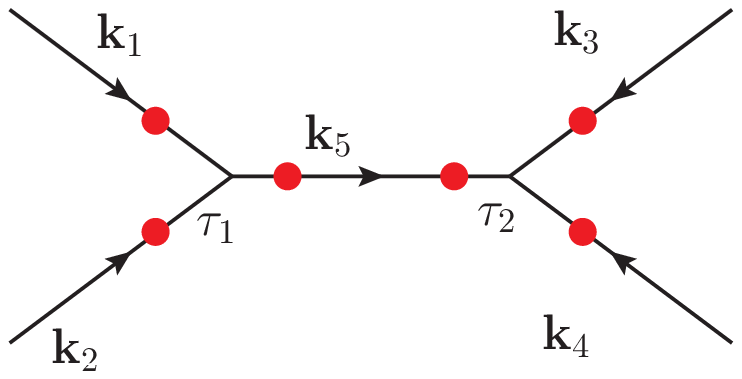}\tabularnewline
$I_{a}$\tabularnewline
\end{tabular} & \begin{tabular}{c}
\begin{tabular}{ccc}
\includegraphics[width=2cm]{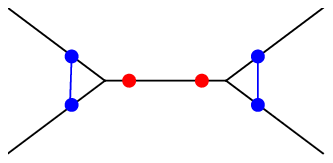} & \includegraphics[width=2cm]{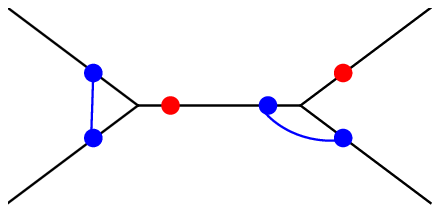} & \includegraphics[width=2cm]{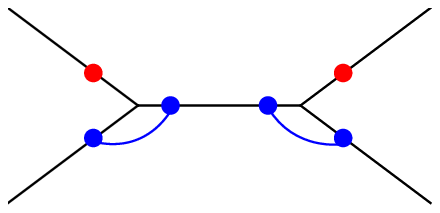}\tabularnewline
$I_{b}^{\left(1\right)}$ & $I_{b}^{\left(2\right)}$ &
$I_{b}^{\left(3\right)}$\tabularnewline
\end{tabular}\tabularnewline
\tabularnewline
\begin{tabular}{cc}
\includegraphics[width=2cm]{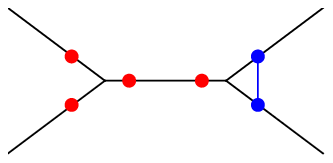} & \includegraphics[width=2cm]{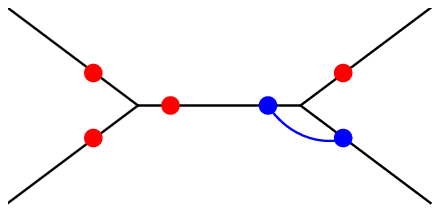}\tabularnewline
$I_{c}^{\left(1\right)}$ & $I_{c}^{\left(2\right)}$\tabularnewline
\end{tabular}\tabularnewline
\end{tabular}\tabularnewline
\end{tabular}
    \label{app_fig_int}
    \caption{Diagrammatic representation of the six basic integrals: $I_a$, $I_b^{(1)}$, $I_b^{(2)}$, $I_b^{(3)}$,
$I_c^{(1)}$ and $I_c^{(1)}$. All the momenta configurations and
$\tau_1$, $\tau_2$ are the same as in $I_a$. A red dot denotes the
temporal derivative, a blue dot denote the spatial derivative or
momentum in Fourier space, where a blue line between two dots
represents the dot product.}
\end{minipage}
\end{figure}

The full expressions for the four-point functions are rather
complicated. In this work, at the leading-order, all contributions
to the four-point functions can be grouped into six basic integrals,
which we denote as $I_a$, $I_b^{(1)}$, $I_b^{(2)}$, $I_b^{(3)}$,
$I_c^{(1)}$ and $I_c^{(1)}$, and their ``conjugate" which we define
as below (see Fig. \ref{app_fig_int}).

From Fig.\ref{app_fig_int}, it is straightforward to read the
expressions for these integrals, we find
    \ea{
    &I_{a}\left(k_{1},k_{2},k_{3},k_{4},k_{5}\right)\\\equiv&-\frac{1}{2H^{2}}\Re\left[\int_{-\infty}^{\tau}d\tau_{1}\int_{-\infty}^{\tau_{1}}d\tau_{2}\frac{1}{\tau_{1}\tau_{2}}\partial_{1}G_{k_{1}}\left(\tau,\tau_{1}\right)\partial_{1}G_{k_{2}}\left(\tau,\tau_{1}\right)\partial_{2}G_{k_{3}}\left(\tau,\tau_{2}\right)\partial_{2}G_{k_{4}}\left(\tau,\tau_{2}\right)\partial_{12}G_{k_{5}}\left(\tau_{1},\tau_{2}\right)\right]\\&+\frac{1}{4H^{2}}\int_{-\infty}^{\tau}d\tau_{1}\int_{-\infty}^{\tau}d\tau_{2}\frac{1}{\tau_{1}\tau_{2}}\partial_{1}G_{k_{1}}\left(\tau_{1},\tau\right)\partial_{1}G_{k_{2}}\left(\tau_{1},\tau\right)\partial_{2}G_{k_{3}}\left(\tau,\tau_{2}\right)\partial_{2}G_{k_{4}}\left(\tau,\tau_{2}\right)\partial_{12}G_{k_{5}}\left(\tau_{1},\tau_{2}\right),
    }
where and in what follows $\partial_{1,2} \equiv
\frac{d}{d\tau_{1,2}}$, $\partial_{12}\equiv \frac{d^2}{d\tau_1
d\tau_2}$ and in this appendix we denote $G_k(\tau_1,\tau_2) =
u_k(\eta_1) u^{\ast}_k(\eta_2)$ with $u_k(\eta) = \frac{i \, H}{
\sqrt{ 2 k^3 }} \lrb{ 1+ i k \eta } e^{-i k \eta} $.
    \ea{
    &I_{b}^{\left(1\right)}\left(\bm{k}_{1},\bm{k}_{2},\bm{k}_{3},\bm{k}_{4},k_{5}\right)\\\equiv&-\frac{1}{2H^{2}}
    \left(\bm{k}_{1}\cdot\bm{k}_{2}\right)\left(\bm{k}_{3}\cdot\bm{k}_{4}\right)
    \Re\left[\int_{-\infty}^{\tau}d\tau_{1}\int_{-\infty}^{\tau_{1}}d\tau_{2}\frac{1}{\tau_{1}\tau_{2}}
    G_{k_{1}}\left(\tau,\tau_{1}\right)G_{k_{2}}\left(\tau,\tau_{1}\right)G_{k_{3}}\left(\tau,\tau_{2}\right)
    G_{k_{4}}\left(\tau,\tau_{2}\right)\partial_{12}G_{k_{5}}\left(\tau_{1},\tau_{2}\right)\right]\\
    & +\frac{1}{4H^{2}}\left(\bm{k}_{1}\cdot\bm{k}_{2}\right)
    \left(\bm{k}_{3}\cdot\bm{k}_{4}\right)\int_{-\infty}^{\tau}d\tau_{1}\int_{-\infty}^{\tau}d\tau_{2}
    \frac{1}{\tau_{1}\tau_{2}}G_{k_{1}}\left(\tau_{1},\tau\right)G_{k_{2}}\left(\tau_{1},\tau\right)G_{k_{3}}\left(\tau,\tau_{2}\right)
    G_{k_{4}}\left(\tau,\tau_{2}\right)\partial_{12}G_{k_{5}}\left(\tau_{1},\tau_{2}\right),
}
    \ea{
    &I_{b}^{\left(2\right)}\left(\bm{k}_{1},\bm{k}_{2},k_{3},\bm{k}_{4},\bm{k}_{5}\right)\\
    \equiv
    &-\frac{1}{2H^{2}}\left(\bm{k}_{1}\cdot\bm{k}_{2}\right)\left(\bm{k}_{5}\cdot\bm{k}_{4}\right)
    \Re\left[\int_{-\infty}^{\tau}d\tau_{1}\int_{-\infty}^{\tau_{1}}d\tau_{2}\frac{1}{\tau_{1}\tau_{2}}
    G_{k_{1}}\left(\tau,\tau_{1}\right)G_{k_{2}}\left(\tau,\tau_{1}\right)\partial_{2}G_{k_{3}}\left(\tau,\tau_{2}\right)
    G_{k_{4}}\left(\tau,\tau_{2}\right)\partial_{1}G_{k_{5}}\left(\tau_{1},\tau_{2}\right)\right]\\
    &+\frac{1}{4H^{2}}\left(\bm{k}_{1}\cdot\bm{k}_{2}\right)\left(\bm{k}_{5}\cdot\bm{k}_{4}\right)
    \int_{-\infty}^{\tau}d\tau_{1}\int_{-\infty}^{\tau}d\tau_{2}\frac{1}{\tau_{1}\tau_{2}}G_{k_{1}}\left(\tau_{1},\tau\right)
    G_{k_{2}}\left(\tau_{1},\tau\right)\partial_{2}G_{k_{3}}\left(\tau,\tau_{2}\right)G_{k_{4}}\left(\tau,\tau_{2}\right)
    \partial_{1}G_{k_{5}}\left(\tau_{1},\tau_{2}\right),
    }
    \ea{
    &I_{b}^{\left(3\right)}\left(k_{1},\bm{k}_{2},k_{3},\bm{k}_{4},\bm{k}_{5}\right)\\
    \equiv&\frac{1}{2H^{2}}\left(\bm{k}_{5}\cdot\bm{k}_{2}\right)\left(\bm{k}_{5}\cdot\bm{k}_{4}\right)
    \Re\left[\int_{-\infty}^{\tau}d\tau_{1}\int_{-\infty}^{\tau_{1}}d\tau_{2}\frac{1}{\tau_{1}\tau_{2}}\partial_{1}
    G_{k_{1}}\left(\tau,\tau_{1}\right)G_{k_{2}}\left(\tau,\tau_{1}\right)\partial_{2}G_{k_{3}}\left(\tau,\tau_{2}\right)
    G_{k_{4}}\left(\tau,\tau_{2}\right)G_{k_{5}}\left(\tau_{1},\tau_{2}\right)\right]\\
    &-\frac{1}{4H^{2}}\left(\bm{k}_{5}\cdot\bm{k}_{2}\right)\left(\bm{k}_{5}\cdot\bm{k}_{4}\right)
    \int_{-\infty}^{\tau}d\tau_{1}\int_{-\infty}^{\tau}d\tau_{2}\frac{1}{\tau_{1}\tau_{2}}\partial_{1}
    G_{k_{1}}\left(\tau_{1},\tau\right)G_{k_{2}}\left(\tau_{1},\tau\right)\partial_{2}G_{k_{3}}\left(\tau,\tau_{2}\right)
    G_{k_{4}}\left(\tau,\tau_{2}\right)G_{k_{5}}\left(\tau_{1},\tau_{2}\right),
    }
and
    \ea{
        &I_{c}^{\left(1\right)}\left(k_{1},k_{2},\bm{k}_{3},\bm{k}_{4},k_{5}\right)\\
        \equiv&\frac{1}{2H^{2}}\left(\bm{k}_{3}\cdot\bm{k}_{4}\right)\Re\int_{-\infty}^{\tau}d\tau_{1}\int_{-\infty}^{\tau_{1}}
        d\tau_{2}\frac{1}{\tau_{1}\tau_{2}}\partial_{1}G_{k_{1}}\left(\tau,\tau_{1}\right)\partial_{1}
        G_{k_{2}}\left(\tau,\tau_{1}\right)G_{k_{3}}\left(\tau,\tau_{2}\right)G_{k_{4}}\left(\tau,\tau_{2}\right)
        \partial_{12}G_{k_{5}}\left(\tau_{1},\tau_{2}\right)\\
        &-\frac{1}{4H^{2}}\left(\bm{k}_{3}\cdot\bm{k}_{4}\right)\int_{-\infty}^{\tau}d\tau_{1}\int_{-\infty}^{\tau}d\tau_{2}
        \frac{1}{\tau_{1}\tau_{2}}\partial_{1}G_{k_{1}}\left(\tau_{1},\tau\right)\partial_{1}G_{k_{2}}\left(\tau_{1},\tau\right)
        G_{k_{3}}\left(\tau,\tau_{2}\right)G_{k_{4}}\left(\tau,\tau_{2}\right)\partial_{12}G_{k_{5}}\left(\tau_{1},\tau_{2}\right),
    }
    \ea{
        &I_{c}^{\left(2\right)}\left(k_{1},k_{2},k_{3},\bm{k}_{4},\bm{k}_{5}\right)\\
        \equiv&\frac{1}{2H^{2}}\left(\bm{k}_{5}\cdot\bm{k}_{4}\right)\Re\int_{-\infty}^{\tau}d\tau_{1}\int_{-\infty}^{\tau_{1}}
        d\tau_{2}\frac{1}{\tau_{1}\tau_{2}}\partial_{1}G_{k_{1}}\left(\tau,\tau_{1}\right)\partial_{1}
        G_{k_{2}}\left(\tau,\tau_{1}\right)\partial_{2}G_{k_{3}}\left(\tau,\tau_{2}\right)G_{k_{4}}\left(\tau,\tau_{2}\right)
        \partial_{1}G_{k_{5}}\left(\tau_{1},\tau_{2}\right)\\
        &-\frac{1}{4H^{2}}\left(\bm{k}_{5}\cdot\bm{k}_{4}\right)\int_{-\infty}^{\tau}d\tau_{1}\int_{-\infty}^{\tau}d\tau_{2}
        \frac{1}{\tau_{1}\tau_{2}}\partial_{1}G_{k_{1}}\left(\tau_{1},\tau\right)\partial_{1}G_{k_{2}}\left(\tau_{1},\tau\right)
        \partial_{2}G_{k_{3}}\left(\tau,\tau_{2}\right)G_{k_{4}}\left(\tau,\tau_{2}\right)\partial_{1}G_{k_{5}}\left(\tau_{1},\tau_{2}\right).
    }

It is useful to introduce the ``conjugate" contributions, defined as
follows. Up to the second-order in perturbation theory, there are
two interaction vertices and thus two temporal integrals with
respect to $\tau_1$ and $\tau_2$ respectively. We call two
contributions (diagrams) are conjugate to each other with exchanging
$\tau_1\leftrightarrow \tau_2$ while keeping all the momenta
relations. Having known the expression for a diagram, it is easy to
write down the integral expression for its conjugate, e.g.
    \ea{
        &\tilde{I}_{a}\left(k_{1},k_{2},k_{3},k_{4},k_{12}\right)\\
        \equiv&-\frac{1}{2H^{2}}\Re\left[\int_{-\infty}^{\tau}d\tau_{1}\int_{-\infty}^{\tau_{1}}d\tau_{2}
        \frac{1}{\tau_{1}\tau_{2}}\partial_{2}G_{k_{1}}\left(\tau,\tau_{2}\right)
        \partial_{2}G_{k_{2}}\left(\tau,\tau_{2}\right)\partial_{1}G_{k_{3}}\left(\tau,\tau_{1}\right)
        \partial_{1}G_{k_{4}}\left(\tau,\tau_{1}\right)\partial_{12}G_{k_{5}}
        \left(\tau_{1},\tau_{2}\right)\right]\\
        &+\frac{1}{4H^{2}}\left[\int_{-\infty}^{\tau}d\tau_{1}\int_{-\infty}^{\tau}d\tau_{2}\frac{1}{\tau_{1}\tau_{2}}
        \partial_{1}G_{k_{1}}\left(\tau_{1},\tau\right)\partial_{1}G_{k_{2}}\left(\tau_{1},\tau\right)\partial_{2}G_{k_{3}}
        \left(\tau,\tau_{2}\right)\partial_{2}G_{k_{4}}\left(\tau,\tau_{2}\right)\partial_{12}G_{k_{5}}\left(\tau_{1},\tau_{2}\right)\right]^{\ast},
    }
where $\ast$ denotes complex conjugate. It is analogous for the
other conjugate integrals, which we do not write here for
simplicity. Moreover, we introduce the combination of a contribution
and its conjugate, e.g.
    \eq{ \mathcal{I}_a\left(k_{1},k_{2},k_{3},k_{4},k_{5}\right)
\equiv [I_a + \tilde{I}_a]
\left(k_{1},k_{2},k_{3},k_{4},k_{5}\right).
    }

Before we evaluate the integrals, it is useful to make it clear
about the smallest set of integrals we need. There are two cases.
For left-right asymmetric diagrams, e.g. $I_b^{(2)}$ (or
$\tilde{I}_b^{(2)}$), we always encounter the combination
$\mathcal{I}_b^{(2)}\equiv I_b^{(2)} + \tilde{I}_b^{(2)}$ rather
than $\tilde{I}_b^{(2)}$ itself.
    While for the left-right symmetric
diagrams, e.g. $I_a$, $\tilde{I}_a$ is simply exchanging
simultaneously $\bm{k}_1 \leftrightarrow \bm{k}_3$ , $\bm{k}_2
\leftrightarrow \bm{k}_4$. Thus, after the 6 permutations (which
specify two momenta associated with $\tau_1$ and other two momenta
associated with $\tau_2$) among the four extra momenta
$\bm{k}_1,\cdots,\bm{k}_4$, the final contribution to the
correlation function from $I_a$ is equal to $\mathcal{I}_a/2$. Thus,
what we really need is the following six basic integrals:
$\mathcal{I}_a$, $\mathcal{I}_b^{(1)}$, $\mathcal{I}_b^{(2)}$,
$\mathcal{I}_b^{(3)}$, $\mathcal{I}_c^{(1)}$ and
$\mathcal{I}_c^{(2)}$.

Now we collect the final results for these integrals, in the limit
of $\tau\rightarrow 0$. We find
    \eq{
        \mathcal{I}_{a}\left(k_{1},k_{2},k_{3},k_{4},k_{5}\right)\equiv\frac{H^{8}k_{5}}{16K^{5}}\left(\prod_{i=1}^{4}\frac{1}{k_{i}}\right)\left[A\left(K_{12},K_{34},k_{5}\right)+\frac{K^{5}}{\left(K_{12}+k_{5}\right)^{3}\left(K_{34}+k_{5}\right)^{3}}\right]
        \,,
    }
with
    \eq{
        A\left(s_{1},s_{2},r\right) \equiv \frac{10s_{2}^{2}+\left(s_{1}+3r\right)\left(4s_{2}+K\right)+6r^{2}}{\left(s_{2}+r\right)^{3}}+\left(s_{1}\leftrightarrow
        s_{2}\right) \,,
    }
where here and in what follows we denote $K_{ij} \equiv k_i + k_j$
and $K\equiv k_1 + k_2 + k_3 +k_4$.
    \ea{
        \mathcal{I}_{b}^{\left(1\right)}\left(\bm{k}_{1},\bm{k}_{2},\bm{k}_{3},\bm{k}_{4},k_{5}\right)
        & = \left(\bm{k}_{1}\cdot\bm{k}_{2}\right)\left(\bm{k}_{3}\cdot\bm{k}_{4}\right)\frac{H^{8}k_{5}}{64K^{5}}
        \left(\prod_{i=1}^{4}\frac{1}{k_{i}^{3}}\right)\\& \times\left[\Gamma \left(K_{12},K_{34};k_{1}k_{2},k_{3}k_{4};J_{12}\right)+\left(12\leftrightarrow34\right)+K^{5}F\left(K_{12},k_{5},k_{1}k_{2}\right)F\left(K_{34},k_{5},k_{3}k_{4}\right)\right].
        }
with $J_{ij} \equiv K_{ij}- k_5$, and
    \ea{
        \Gamma(s_1,s_2;q_1,q_2,t)
        \equiv&\frac{1}{\left(K-t\right)^{3}}\left\{K^{6}+K^{5}\left(-2t+s_{1}
        +2s_{2}\right)\right.\\&+K^{4}\left[t\left(t-2s_{1}\right)+3\left(-t+s_{1}\right)s_{2}
        +2q_{1}+6q_{2}\right]\\&+K^{3}\left[t^{2}\left(s_{1}+s_{2}\right)+8s_{2}q_{1}
        +12s_{1}q_{2}-t\left(5s_{1}s_{2}+4q_{1}+6q_{2}\right)\right]\\&+K^{2}\left[2t\left(ts_{1}s_{2}+\left(t-7s_{2}\right)q_{1}\right)+2\left(t^{2}-8ts_{1}+20q_{1}\right)q_{2}\right]\\&\left.+6Kt\left[ts_{2}q_{1}+\left(ts_{1}-10q_{1}\right)q_{2}\right]+24t^{2}q_{1}q_{2}\right\}
        \,,
    }
and
    \eq{
        F\left(s,t,q\right) \equiv \frac{2s^{2}+2q+3st+t^{2}}{\left(s+t\right)^{3}}.
    }
    \ea{
        &\mathcal{I}_{b}^{\left(2\right)}\left(\bm{k}_{1},\bm{k}_{2},k_{3},\bm{k}_{4},\bm{k}_{5}\right)\\
        \equiv&\left(\bm{k}_{1}\cdot\bm{k}_{2}\right)\left(\bm{k}_{4}\cdot\bm{k}_{5}\right)\left(\frac{k_{3}}{k_{5}}\right)^{2}
        \frac{H^{8}k_{5}}{64K^{5}}\left(\prod_{i=1}^{4}\frac{1}{k_{i}^{3}}\right)\\&\times
        \left[{\Gamma}\left(K_{12},K_{45};k_{1}k_{2},k_{4}k_{5};J_{12}\right)+{\Gamma}\left(\bar{K}_{45},K_{12};-k_{4}k_{5},k_{1}k_{2};J_{34}\right)+K^{5}F\left(K_{12},k_{5},k_{1}k_{2}\right)F\left(K_{45},k_{3},k_{4}k_{5}\right)\right]
        \,,
        }
where $\bar{K}_{ij} \equiv k_i - k_j$.
    \ea{
        &\mathcal{I}_{b}^{\left(3\right)}\left(k_{1},\bm{k}_{2},k_{3},\bm{k}_{4},\bm{k}_{5}\right)\\
        \equiv&\left(-\bm{k}_{2}\cdot\bm{k}_{5}\right)\left(\bm{k}_{4}\cdot\bm{k}_{5}\right)
        \frac{k_{1}^{2}k_{3}^{2}}{k_{5}^3}
        \frac{H^{8}}{64K^{5}}\left(\prod_{i=1}^{4}\frac{1}{k_{i}^{3}}\right)\\&\times\left[
        {\Gamma}\left(\bar{K}_{25},K_{45};-k_{2}k_{5},k_{4}k_{5};J_{12}\right)
        +{\Gamma}\left(\bar{K}_{45},K_{25};-k_{4}k_{5},k_{2}k_{5};
        J_{34}\right)+K^{5}F\left(K_{25},k_{1},k_{2}k_{5}\right)F\left(K_{45},k_{3},k_{4}k_{5}\right)\right]
        \,.}
And
    \ea{
       &\mathcal{I}_{c}^{\left(1\right)}\left(k_{1},k_{2},\bm{k}_{3},\bm{k}_{4},k_{5}\right)\\
       \equiv&\left(\bm{k}_{3}\cdot\bm{k}_{4}\right)\frac{H^{8}}{32K^{5}}\frac{k_{5}}{k_{1}k_{2}k_{3}^{3}k_{4}^{3}}
       \left[C\left(K_{34},k_{3}k_{4},J_{12}\right)+\bar{C}\left(K_{34},k_{3}k_{4},J_{34}\right)
       +\frac{K^{5}F\left(K_{34},k_{5},k_{3}k_{4}\right)}{\left(K_{12}+k_{5}\right)^{3}}\right]
       \,,}
    \ea{
        &\mathcal{I}_{c}^{\left(2\right)}\left(k_{1},k_{2},k_{3},\bm{k}_{4},\bm{k}_{5}\right)\\\equiv&\left(\bm{k}_{5}\cdot\bm{k}_{4}\right)\frac{H^{8}}{32K^{5}}\frac{1}{k_{1}k_{2}k_{3}k_{4}^{3}k_{5}}\left[C\left(K_{45},k_{4}k_{5},J_{12}\right)+\bar{C}\left(\bar{K}_{45},-k_{4}k_{5},J_{34}\right)+\frac{K^{5}F\left(K_{45},k_{3},k_{4}k_{5}\right)}{\left(K_{12}+\tilde{k}_{5}\right)^{3}}\right]
        \,,}
with
    \ea{
        C\left(s,q,t\right) &\equiv \frac{K\left(K-t\right)\left[-t\left(K+3s\right)
        +K\left(K+4s\right)\right]+2\left(10K^{2}-15Kt+6t^{2}\right)q}{\left(K-t\right)^{3}},\\
         \bar{C}\left(s,q,t\right)&\equiv
         \frac{K\left(3K^{2}\left(K+2s\right)+t^{2}\left(K+3s\right)-Kt\left(3K+8s\right)\right)+2\left(10K^{2}-15Kt+6t^{2}\right)q}{\left(K-t\right)^{3}}
         \,.
    }


\end{document}